\documentclass{article}
\usepackage{spconf,amsmath,graphicx,subfigure,multirow,diagbox,color,fontenc,amsfonts}
\usepackage{booktabs}

\title{CAN WE TRUST DEEP SPEECH PRIOR?}
\name{Ying Shi$^1$, Haolin Chen$^2$, Zhiyuan Tang$^2$, Lantian Li$^{2*}$, Dong Wang$^{2*}$, Jiqing Han$^{1*}$
\thanks{
This work was supported by the National Natural Science Foundation of China (NSFC) under the project No.~61633013.
L.L. (lilt@cslt.org), D.W. (wangdong99@mails.tsinghua.edu.cn) and J.H. (jqhan@hit.edu.cn) are the corresponding authors.
}
}
\address{Harbin Institute of Technology, Harbin, China\\
Center for Speech and Language Technologies, Tsinghua University, China
}
\begin{document}
%\ninept
%
\maketitle
\begin{abstract}

 Recently, speech enhancement (SE) based on deep speech prior has attracted much attention,
 such as the variational auto-encoder with non-negative matrix factorization (VAE-NMF) architecture. Compared to conventional approaches that represent clean speech by shallow models such as Gaussians
 with a low-rank covariance, the new approach employs deep generative models to represent the clean speech, which
 often provides a better prior. Despite the clear advantage in theory, we argue that deep priors must be used
 with much caution, since the likelihood produced by a deep generative model does not always coincide with the speech quality.
 We designed a comprehensive study on this issue and demonstrated that based on deep speech priors,
 a reasonable SE performance can be achieved, but the results might be suboptimal.
 A careful analysis showed that this problem is deeply rooted in the disharmony between the flexibility of
 deep generative models and the nature of the maximum-likelihood (ML) training.
 \end{abstract}

\section{introduction}

Single-channel speech enhancement (SE) aims to restore the quality and possibly also the intelligibility of noise corrupted speech~\cite{loizou2013speech,rehr2019robust}.
Most modern SE methods are based on the maximum-likelihood (ML) statistical framework.
The most popular statistical model is the complex Gaussian on the short-time Fourier transform (STFT) coefficients~\cite{brillinger2001time}, although
some super-Gaussian models have been experimented~\cite{martin2005speech}.
The choice for Gaussian is not only theoretically sound (e.g., justified by the central limit theorem~\cite{pearlman1978source}),
but also computational attractive. Specifically,
by assuming that clean speech and noise spectra are both zero-mean Gaussians and are independent and additive,
the ML inference leads to the spectra substraction approach~\cite{boll1979suppression,mcaulay1980speech} or
the Wiener filter~\cite{vary2006digital,bando2018statistical}, depending on different model assumptions.

The ML-based statistical SE approach relies on the covariance estimation of the Gaussians for clean speech and noise.
Non-negative matrix factorization (NMF)~\cite{lee1999learning} is perhaps the most popular variance model, which
assumes that the variances at different time-frequency (T-F) bins possess a low-rank structure.
Although elegant in theory, the low-rank covariance model may be too simple to represent
complex signals such as speech.
Recently, several authors have proposed to use deep generative models to represent speech. For example,
the VAE-NMF architecture~\cite{bando2018statistical,leglaive2018variance,leglaive2020recursive,li2019fast} uses
a variational auto-encoder (VAE) to model the covariance of the complex Gaussian for speech signal, while the noise model is
still based on NMF. Subakan et al.~\cite{ganspe} models the power spectrum of speech signals as a Poisson distribution and trains a deep neural net
to predict the variance of the distribution. The central idea of these studies is to learn a powerful prior model for clean speech,
which we call a deep speech prior.
In theory, this deep prior can provide very rich knowledge for clean speech, hence a powerful enhancement.
This has been demonstrated in~\cite{bando2018statistical,leglaive2018variance}.

In this paper, we argue that caution must be given to use the deep speech prior models.
The argument is based on a key observation that a high likelihood measured by a deep generative model does not mean
that the sample is similar to the training data, which is clean speech in our case. This means that
the likelihood produced by deep speech prior models does not always reflect the quality of the speech,
leading to suboptimal results with the ML-based statistical SE framework.
We will design a serial of experiments to investigate this issue.
In particularly, our analysis is based on normalization flow (NF)~\cite{Ryan2019Waveglow}, a typical deep
generative model that can measure the likelihood accurately.
Our results show that reasonable SE results can be obtained with deep speech priors,
at least when the noise can be well modeled. However, there is a potential risk that the inference converges to
unrealistic speech, due to the unreliable likelihood produced by the deep speech prior models.
We attribute this problem to the disharmony between the flexibility of
deep generative models and the nature of the ML training criterion.

The rest of the paper is organized as follows. Section 2 revisits the ML-based statistical SE framework,
and Section 3 reviews the deep generative models.
The flow-based deep speech prior model and the inference algorithm based on gradient descent (GD) are
presented in Section 4. Experiments and analysis are presented in Section 5, and
Section 6 concludes the paper.

\section{Revisit statistical speech enhancement}

\subsection{ML-based statistical SE framework}

Let $\mathbf{x}_t$ denote a segment of clean speech of length $M$, and $\mathbf{n}_t$ denote a segment of noise signal in the same length.
Suppose the noise is additive, the noisy speech $\mathbf{y}_t$ can be written by:

\begin{equation}
\label{eq:xn}
\mathbf{y}_t = \mathbf{x}_t + \mathbf{n}_t.
\end{equation}

\noindent The clean speech can be estimated from the noisy speech following the ML principle\footnote{This is essentially the
\emph{maximum a posterior} (MAP) principle, though we regard it as a generalized ML, considering that it maximizes the conditional likelihood.}:

\begin{eqnarray}
\hat{\mathbf{x}}_t &=& \arg\max_{\mathbf{x}_t} p(\mathbf{x}_t|\mathbf{y}_t) \nonumber \\
&=& \arg\max_{\mathbf{x}_t} p(\mathbf{x}_t) p(\mathbf{y}_t|\mathbf{x}_t) \nonumber \\
&=& \arg\max_{\mathbf{x}_t} p(\mathbf{x}_t) p(\mathbf{n}_t) \label{eq:time-map},
\end{eqnarray}
\noindent where we have assumed $\mathbf{x}_t$ and $\mathbf{n}_t$ are independent.

Let $\pmb{s}^x_t, \pmb{s}^n_t, \pmb{s}^y_t$ denote the complex spectrum of $\mathbf{x}_t,\mathbf{n}_t,\mathbf{y}_t$ respectively.
The superposition of Eq.(\ref{eq:xn}) still holds since the Fourier transform is a linear operator.
\begin{equation}
\label{eq:xn-s}
\pmb{s}^y_t = \pmb{s}^x_t + \pmb{s}^n_t.
\end{equation}

\noindent The ML estimation for $\pmb{s}^x_t$ has a similar form as in Eq.(\ref{eq:time-map}):
\begin{equation}
\label{eq:post}
\hat{\pmb{s}}^x_t  =\arg\max_{\pmb{s}^x_t} p(\pmb{s}^x_t) p(\pmb{s}^n_t).
\end{equation}

\noindent If $p(\pmb{s}^x_t)$ and $p(\pmb{s}^n_t)$ are all complex Gaussians and the covariances are
diagonal, i.e.,

\[
p(s^x_{tf}) = \mathcal{N}_C (0, \sigma_{tf}^x)
\]
\[
p(s^n_{tf}) = \mathcal{N}_C (0, \sigma_{tf}^n),
\]

\noindent the posterior $p(\pmb{s}^x_{t}|\pmb{s}^y_t)$ can be represented in the following element-wised form:

\[
p(s^x_{tf}|s^y_{tf}) = \mathcal{N}_C (\frac{\sigma^x_{tf}}{\sigma^x_{tf} + \sigma^n_{xf}} y_{tf}, \frac{\sigma^n_{tf} \sigma^x_{tf}}{\sigma^n_{tf} + \sigma^x_{tf}}).
\]
\noindent Therefore, the ML estimation for $\pmb{s}^x_t$ can be written by:

\begin{equation}
\label{eq:wf}
\hat{s}^x_{tf} = \frac{\sigma^x_{tf}}{\sigma^x_{tf} + \sigma^n_{tf}} y_{tf}.
\end{equation}
\noindent Note that this is the form of Wiener filter~\cite{vary2006digital,bando2018statistical}.

\subsection{NMF prior model}

The paramount task of the statistical SE approach is to estimate the variance $\sigma^x_{tf}$ and $\sigma^n_{tf}$
for each T-F bin. Note that they correspond to the power spectral density (PSD) of clean speech and noise, respectively.
Traditional algorithms focus on the noise PSD, either heuristically (e.g., by voice activity detection~\cite{sohn1998voice,mcaulay1980speech})
or principally (e.g., by speech present probability~\cite{hirsch1995noise,gerkmann2011noise,chinaev2016noise}). Since this
estimation heavily relies on the assumption that the noise is time-invariant (at least change slowly), it cannot
deal with dynamic noises. A principle solution is to learn the patterns of the
speech and noise \emph{in prior}, and then infer the clean speech based on these prior models.
For example, hidden Markov models (HMMs)~\cite{ephraim1992bayesian} or Gaussian mixture models (GMM)~\cite{hao2009speech} were
used as the prior models in early research.

NMF~\cite{lee1999learning} is perhaps the most popular prior model. Initial application of NMF to SE is largely heuristic, which distributes the PSD of noisy speech to
PSDs of clean speech and noise, e.g., $\mathbf{V}_y \approx \mathbf{V}_x + \mathbf{V}_n$,
where each PSD is in the form of $\mathbf{V} \approx \mathbf{W}\mathbf{H}$,  where
$\mathbf{W}_{0,+}^{K \times L}$ represents $L$ bases, and $\mathbf{H} \in R_{0,+}^{L \times T}$ represents the activations
on the bases~\cite{smaragdis2006convolutive,wang2012online}. Usually $L << K$, and so NMF represents a low-rank decomposition for the PSD.
%To distinguish clean speech and noise,
%the basis matrices for the clean speech and/or the noise need to be pre-trained, so they are prior models.
F{\'e}votte et al.~\cite{fevotte2009nonnegative} demonstrated that if the spectrum follows a complex Gaussian and the
variances on all the T-F bins are assumed to be in form of $\pmb{\Sigma}=\mathbf{WH}$,
the ML estimation with respect to $\mathbf{W}$ and $\mathbf{H}$ is equivalent to the NMF
on $\mathbf{V}$ with the Itakura-Saito (IS) divergence.
Due to this equivalence, the ML estimation for the variances (equivalently, the PSD)
of clean speech and noise can be conducted by IS-based NMF.

%Note that the NMF-based SE approach can be either supervised~\cite{smaragdis2006convolutive} or semi-supervised~\cite{mohammed2017statistical}.
%In the supervised setting, the basis matrices for both the clean speech and noise are pre-trained, while
%in the semi-supervised setting, the basis matrix for only clean speech is pre-trained, leaving the noise bases estimated as part of the inference.
%A particular advantage of the semi-supervised approach is that it can deal with unseen noise,
%with the only request that the covariance of the noise is low-rank.

%MMSE estimator for amplitude, by assuming the amplitude of speech is Gaussian~enhancemtn\cite{ephraim1984speech}
%However, experiments that have been conducted to measure thePDF of clean speechspectral coefficients showed that the Gaussian assumption may not be appropriate~\cite{martin2005speech}.
%Most STFT-based filtering schemes assumethat the speech signal is corrupted by additive noise which yields the noisy signal. This assumption is often used because it is well motivated by the physical properties ofsound [43].

\section{Deep speech prior: potential and risk}

\subsection{Deep speech prior}

A key advantage of the NMF-based SE approach is that the activation
matrix $\mathbf{H}$ is estimated on-the-fly during inference, thus compensating for the variation on the magnitude of the signal.
Moreover, since the distributions of both clean speech and noise are Gaussian, the inference is simply in the form of Wiener filtering.
The disadvantage is that the low-rank form of the variances may prevent it from representing complex signals, e.g., speech.

Recently, deep generative models have been extensively studied,
including the auto-regressive model~\cite{van2016conditional,oord2016wavenet},  variational auto-encoder (VAE)~\cite{vaekingma},
normalization flow~\cite{dinh2014nice,dinh2016density,kingma2018glow}, and generative adversarial network (GAN)~\cite{goodfellow2014generative}.
A key advantage of the deep generative models is that they can represent complex natural data, by either density
estimation or sampling, therefore suitable for modeling speech signals.

Several groups have proposed to use deep generative models to present clean speech in SE, which
we call deep speech prior models. Among the proposals, VAE-NMF is in particular
attractive~\cite{bando2018statistical,leglaive2018variance,leglaive2020recursive,li2019fast}.
In this architecture, VAE is used to model the covariance of the complex Gaussian for clean speech, and the noise model is
still based on NMF.
Since both the models are still Gaussian, the inference remains in the form of Wiener filtering.
%though the estimation for the covariances of the Gaussians can never be performed by NMF.

%However, because the covariances of the two components are not consistent, the inference has to resort to
%some general algorithms (e.g., variational methods or sampling), rather than the simple NMF with IS divergence.

\subsection{Pitfall of deep speech prior}

Although interesting, some recent research observations cast a shadow on
the employment of deep speech prior models.
For example, Nalisnick et al.~\cite{nalisnick2018deep} and Choi et al.~\cite{choi2018generative} concurrently
found that a normalization flow may assign out-of-domain data
a high likelihood, even higher than that of the training data. They also found that the same problem occurred
in VAE and PixelCNN.
This is very weird at the first glance, as all the models are trained with the ML criterion, and
therefore should assign the highest likelihood to the training data.

\begin{figure}[h]
	\centering
    \includegraphics[width=\linewidth]{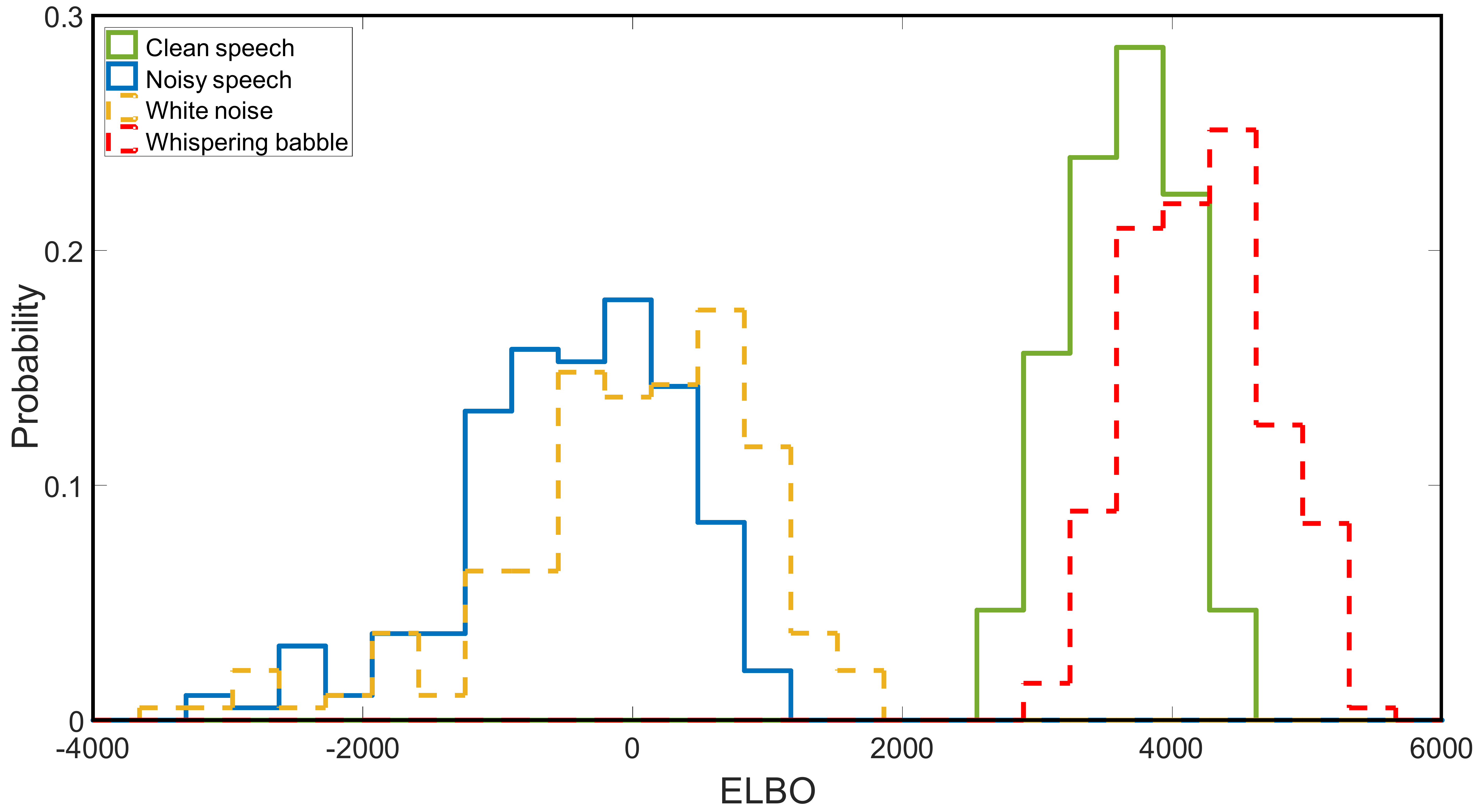}
	\caption{Histogram of ELBO for clean speech (green solid stairs), noisy speech (blue solid stairs), white noise (orange dashed stairs),
whispering babble (red dashed stairs), computed from a VAE trained with clean speech.}\
	\label{fig:vae}
\end{figure}

To investigate if the same problems exists with the deep speech prior, we trained a VAE-NMF model using the code released\footnote{https://github.com/sleglaive/MLSP-2018} by Leglavie et al.~\cite{leglaive2018variance},
and plotted the histogram of the likelihood with different types of speech data.
Since the likelihood cannot be exactly obtained with VAE, we instead use the evidence lower bound (ELBO). Note that the VAE training
maximizes ELBO, so the ELBO values on the training data (clean speech) are expected to be the highest.
The results are shown in Fig.\ref{fig:vae}. It can be seen that the clean speech and noisy speech are well separated, which is nice;
however, the white noise seems obtain higher ELBO than the noisy speech, which is not consistent with the intuition.
More seriously, the whispering babble obtains even higher ELBO than the clean speech.

The disharmony between likelihood (or ELBO) and speech quality (more precisely, the resemblance to the training data) is a disaster
for the ML-based statistical SE, as the ML principle does not optimize the speech quality, and may result in unrealistic speech.

\section{Deep speech prior with normalization flow}

In this paper, we will conduct a comprehensive study for the deep speech prior.
To this purpose, we made several decisions as shown below.
All these decisions help us isolate the issue with the deep speech prior model.

\begin{itemize}
 \item We use normalization flow (NF) rather than VAE as the speech prior model. The NF model can measure the likelihood exactly, rather
 than an approximation (i.e., ELBO) as with VAE.
 Specifically, we choose the  WaveGlow~\cite{Ryan2019Waveglow} model, which has shown great potential in speech synthesis and speech conversion.

 \item We build the SE model in the time domain, and model the speech prior $p(\textbf{x}_t)$ directly, rather than assume a Gaussian and then
 model the covariance (as in the VAE-NMF approach). This eliminates all the assumptions/approximations made when transferring to the frequency-domain and
 conducting modeling there.

 \item We constrain our study with white noise. By white noise, the distribution in the time domain is a Gaussian and can be modeled precisely. This will
 exclude potential problems caused by inaccurate noise modeling, and so fully expose the problems caused by the deep speech prior model.

 \item The inference is conducted based on the original ML principle shown in Eq.(\ref{eq:time-map}), based on the gradient descent (GD) algorithm.
  Therefore, it does not rely on any specific assumption/approximation, and so will neither conceal any problem of the model nor introduce nonexisting problems.

\end{itemize}

We will start from reviewing the normalization flow model, and then describe the SE-approach with a flow-based deep speech prior model.

\subsection{Revisit normalization flow and WaveGlow}

Normalization flow (NF)~\cite{dinh2014nice,dinh2016density,kingma2018glow,papamakarios2019normalizing} is a typical deep generative model.
It employs an invertible transformation $f^{-1}(\mathbf{x})$ that transforms a complex distribution $p(\mathbf{x})$ to a simple
Gaussian. Formally, $\mathbf{z}=f^{-1}(\mathbf{x})$
where $p(\mathbf{z})$ is a Gaussian, i.e.,  ${\mathbf{z}} \sim N({\mathbf{z}};0,{I})$. According to the principle of distribution transformation, we have:

\begin{equation}
\label{eq:flow-x}
 p(\mathbf{x}) = p(\mathbf{z})|\det \frac{\partial f^{-1}} {\partial \mathbf{x}}|,
\end{equation}

\noindent where $\det \frac{\partial f^{-1}} {\partial \mathbf{x}}$ is the determinant of the Jacobian of $f^{-1}$ at $\mathbf{x}$.
The model is trained with the ML criterion, where the objective is as follows:

\begin{equation}
 \sum_i \log p(\mathbf{x}_i) = \sum_i \log p(\mathbf{z}_i) + \sum_i \log |\det \frac{\partial f^{-1}} {\partial \mathbf{x}_i}|.
 \label{equ:logp}
\end{equation}

\noindent After the training converges, the likelihood $p(\mathbf{x})$ can be computed according to Eq.(\ref{eq:flow-x}).
Papamakarios's review paper and the references therein are good resource for the NF model~\cite{papamakarios2019normalizing}.

WaveGlow is a special NF model~\cite{Ryan2019Waveglow}. The structure is inherited from Glow~\cite{kingma2018glow},
with modification to suit for processing speech signals. In our experiment, we reused the WaveGlow architecture
presented in~\cite{Ryan2019Waveglow}. The input is a window of 80 speech samples in the time domain, and the output
is the latent variable with the same dimension as the input.
The entire flow involves 12 blocks, and each consists of two components:
an invertible 1 $\times$ 1 convolution layer and an affine coupling layer.
The output of the latent variable $\mathbf{z}$ is arranged in a multi-scale way, i.e.,
20 dimensions are output every 4 flow blocks.
Different from the original WaveGlow model, we do not need the
conditional input as our focus is in the marginal distribution of $p(\mathbf{x})$.
Readers can refer to Ryan's paper~\cite{Ryan2019Waveglow} for more details of the WaveGlow structure.

\subsection{Speech enhancement with flow-based prior}
\label{sec:mlecp}

A nice property of the NF model is that the exact likelihood $p(\mathbf{x})$ can be computed via Eq.(\ref{eq:flow-x}),
without any assumption on the distribution (as in NMF) or any approximation on the computation (as in VAE). This
permits us pursuing the ML inference for the clean speech by optimizing Eq.(\ref{eq:post}) directly, by using any
numerical optimization algorithm, such as gradient descent (GD).

In theory, the flow model can be used to model the prior for both the clean speech and the noise signal, and the
inference based on Eq.(\ref{eq:post}) is theoretically optimal if
the prior models produce accurate likelihood for clean speech and noise.
However, as we have discussed, the likelihood produced by
deep generative models, including the WaveGlow model we used, is suspicious. To simplify the
investigation, we use the flow model to represent the clean speech only, and constrain the noise to be Gaussian.
Therefore, the ML-based inference can be conducted by optimizing the following objective with respect to $\mathbf{x}_t$:

\begin{eqnarray}
  \log p(\mathbf{x}_t|\mathbf{y}_t) &\propto& \log p(\mathbf{x}_t) + \log p(\mathbf{n}_t) \nonumber \\
                   &=& \log N(\mathbf{z_t}; \mathbf{0} , \mathbf{I}) + \log |\det \frac{\partial f^{-1}} {\partial \mathbf{x}_t}|
                   \nonumber\\
                   && + \log N(\mathbf{y}_t - \mathbf{x}_t; \mathbf{0}, \sigma \mathbf{I} )
  \label{eq:Flow-Gauss}
\end{eqnarray}

\noindent where $\mathbf{z_t}=f^{-1}(\mathbf{x_t})$ and $f^{-1}$ is the flow model.
$\sigma$ is the variance of the noise. Since all the terms in the above objective can be computed
easily, the optimization can be easily conducted by GD.
Note that the Gaussian noise is just the white noise which is ubiquitous in ambient and most destructive for speech signal.

It should be highlighted that both the noise model and the inference are accurate in our SE architecture, which means that any
unexpected behaviors should be attributed to the speech prior model $p(\mathbf{x}_t)$. We therefore expose the
speech prior model by this specific design.

%The above optimization is MAP optimal, which will be denoted by the \emph{Flow-Gauss} approach.
%We can also derive the clean speech by optimizing the speech prior only.
%That is, start from the noisy speech $\mathbf{y}_t$ and optimize $\log p(\mathbf{x})$ by GD,
%and stop the process until the likelihood is improved to a reasonable value as clean speech in the training data.
%This approach, which combines prior optimization and early stop, will be denoted by \emph{Flow-prior}.

\section{Experiments and analysis}

\subsection{Data}
\label{sec:data}

We use the TIMIT dataset~\cite{garofolo1993timit} in our experiments.
For the training set, 4620 utterances were selected from 462 speakers (136 females and 326 males), and the total amount of speech was 4 hours.
It was used to train the prior models (NMF, VAE and WaveGlow).
For the test set, 192 utterances were selected from 25 speakers (8 females and 16 males),
and then corrupted by white noise at 4 different signal-to-noise ratios (SNRs) (-6, 0, 6 and 9~dB).
Perceptual evaluation of speech quality (PESQ)~\cite{pesq}, short-time objective intelligibility (STOI)~\cite{stoi},
and frequency-weighted segmental SNR (fwSNRseg)~\cite{bssevaluation} were used as metrics to evaluate the SE performance.

%For NMF, the noise training data are 2 hours' white noise with various energy.

%\subsubsection{Metrics}
%
%Three metrics are used to evaluate the SE performance:
%
%\begin{itemize}
%\item Perceptual evaluation of speech quality (PESQ)~\cite{pesq}: A metric that considers the perceptual impact when measuring the difference
%between enhanced signal and original signal.
%\item Short-time objective intelligibility (STOI)~\cite{stoi}: A metric that pays more attention on the intelligibility of short-time speech segments.
%\item Frequency-weighted segmental SNR (fwSNRseg)~\cite{bssevaluation}: An SNR measure averaged over different frequency bands.
%\end{itemize}

\subsection{Models}

We test three SE models: NMF, VAE-NMF and Flow-Gauss. The features and settings of these models are
described as follows:

\begin{itemize}

\item \textbf{NMF}: A supervised frequency-domain model where both clean speech and noise are modeled by a complex Gaussian whose
variance is modeled by NMF.
The model is based on 257-dimensional spectrum. The number of
bases is empirically set to 40 for both speech and noise NMFs.
The speech bases are trained with the TIMIT training set,
and the noise bases are trained with 2 hours of white noise sampled with different levels of power.
The inference is based on the IS-based NMF.

\item \textbf{VAE-NMF}: A supervised frequency-domain model, built on 257-dimensional spectrum.
The prior model for the clean speech is a complex Gaussian whose covariance is modeled by a VAE,
and the noise model is a complex Gaussian whose covariance is modeled by an NMF.
The VAE model is constructed by using the source code provided by Leglaive~\cite{leglaive2018variance},
and trained with the TIMIT training set. The base matrix $W$ of the NMF is trained with 2 hours of
white noise with different levels of variances,
and the activation $H$ is derived on-the-fly during the inference.
The inference is based on Bayesian inference~\cite{leglaive2018variance}.
%{\color{red} [better to have NMF supervised, i.e., learn the noise bases before hand and do not change it
%during the inference. By this setting, we will see all the models are supervised.]}

%\item \textbf{Flow-prior}: The SE approach based on deep speech prior only. The speech prior model is  WaveGlow, built on the time-domain.
%We use the source code provided by~\cite{Ryan2019Waveglow}. The input to the model is a  window of $80$ samples,
%while the condition is based on $xx$ samples preceding to the present window. The model was trained on a
%single GPU using randomly chosen clips of 16,000 samples for 200,000 iterations with
%the Adam optimizer, with a batch size of 4 and a learning rate of $1 \times 10^{-4}$.
%The inference is based on GD, using the ADAM optimizer with a learning rate of $5 \times 10^{-4}$.
%The optimization will be stop when the likelihood is as high as a normal clean speech, for which
%we set a threshold  $-7.0$ in our experiments.

\item \textbf{Flow-Gauss}: The time-domain approach proposed in Section~\ref{sec:mlecp}.
The speech prior model is a WaveGlow and the noise model is a Gaussian $N(\mathbf{0}, \sigma \mathbf{I})$.
The WaveGlow is implemented using the source code provided by~\cite{Ryan2019Waveglow},
where the input is a window of 80 samples.
The Adam optimizer is used to train the model, with the batch size set to 4 and the learning rate set to 0.0001.
The inference is based on GD, using the Adam optimizer with the learning rate of 0.0005.
The optimization will be stopped if the likelihood is as high as the clean speech,
for which we set a threshold -7.0 in our experiments.

We test two settings for the variance $\sigma$ of the Gaussian noise:
(1) \textbf{Flow-Gauss (Est)}, where $\sigma$ is manually set to be
100, 200, 2000, 4000 for the SNR conditions -6, 0, 6 and 9 dB respectively. Note that for each SNR condition,
$\sigma$ is different for different utterances due to their different speech magnitudes,
so the setting for $\sigma$ here is an estimated value for the strength of the noise.
(2) \textbf{Flow-Gauss (Opt)}, where the true $\sigma$ is set for each utterance.
Although this is not a practical scenario, the oracle setting for the noise model
is useful for us to investigate the problem of the speech prior model.

%Again, the inference is based on GD, using the Adam optimizer with a learning rate of 0.0001.

%\item \textbf{Flow-Prior}: The SE approach based on deep speech prior only.
%Since the noise model is simply a Gaussian, it is natural to conjecture that the behavior of the flow-based
%SE model is mostly determined by the speech prior. As a reference, we design a purely prior-based approach
%that derive the clean speech by maximizing the speech prior $p(\textbf{z})$.
%The optimization will be stop when the likelihood is as high as a normal clean speech, for which
%we set a threshold $-7.0$ in our experiments. This early stop scheme aims to find the clean speech
%that is a close neighbour of the noisy speech.

\end{itemize}

\subsection{Basic results}

\begin{table*}[htb!]
  \caption{Basic results with different SE models.}
  \label{tab:res}
  \centering
  \scalebox{0.86}{
  \begin{tabular}{l|ccc|ccc|ccc|ccc}
   \cmidrule(r){1-13}
   \multirow{2}{*}{Models} & \multicolumn{3}{c|}{-6dB} & \multicolumn{3}{c|}{0dB} & \multicolumn{3}{c|}{6dB} & \multicolumn{3}{c}{9dB}\\
                           \cmidrule(r){2-4}           \cmidrule(r){5-7}          \cmidrule(r){8-10}         \cmidrule(r){11-13}
                           & PESQ & STOI & fwSNRseg    & PESQ & STOI & fwSNRseg   & PESQ & STOI & fwSNRseg   & PESQ & STOI & fwSNRseg  \\
    \cmidrule(r){1-1}      \cmidrule(r){2-4}           \cmidrule(r){5-7}          \cmidrule(r){8-10}         \cmidrule(r){11-13}
      Noisy                & 1.032 & 0.582 & 4.068     & 1.046 & 0.722 & 4.816    & 1.098 & 0.849 & 6.323    & 1.161 & 0.898 & 7.446 \\
    \cmidrule(r){1-1}      \cmidrule(r){2-4}           \cmidrule(r){5-7}          \cmidrule(r){8-10}         \cmidrule(r){11-13}
      NMF                  & 1.144 & 0.592 & 4.202     & 1.307 & 0.733 & 5.799    & 1.523 & 0.830 & 7.685    & 1.647 & 0.865 & 8.742 \\
    \cmidrule(r){1-1}      \cmidrule(r){2-4}           \cmidrule(r){5-7}          \cmidrule(r){8-10}         \cmidrule(r){11-13}
      VAE-NMF              & \textbf{1.211} & \textbf{0.668} & 3.307 & \textbf{1.319} & \textbf{0.800} & 5.590 & 1.521 & 0.887 & 8.776 & 1.738 & 0.917 & 10.203 \\
    \cmidrule(r){1-1}      \cmidrule(r){2-4}           \cmidrule(r){5-7}          \cmidrule(r){8-10}         \cmidrule(r){11-13}
      Flow-Gauss (Est)     & 1.096 & 0.637 & 5.160     & 1.301 & 0.791 & 7.627    & 1.693 & 0.894 & 10.585   & 1.920 & 0.925 & 12.044 \\
      Flow-Gauss (Opt)     & 1.094 & 0.647 & {\textbf{5.215}} & 1.304 & 0.799 & \textbf{7.772} & \textbf{1.718} & \textbf{0.898} & \textbf{10.796} & \textbf{1.977} & \textbf{0.929} & \textbf{12.388} \\
    \cmidrule(r){1-13}
  \end{tabular}}
\end{table*}

The basic SE results with different models are shown in Table~\ref{tab:res}.
It can be seen that all the SE methods obtained noticeable
performance gains in terms of the three metrics. The VAE-NMF approach generally outperforms the NMF approach, demonstrating
that the deep speech prior model is more powerful. Moreover, the flow-based approach obtained comparable performance as VAE-NMF,
even better in terms of fwSNRseg. In high-SNR conditions, the flow-based models seem more superior.

\subsection{Analysis 1: Maximizing likelihood does not maximize quality}

\begin{figure}[h]
	\centering
	\includegraphics[width=\linewidth]{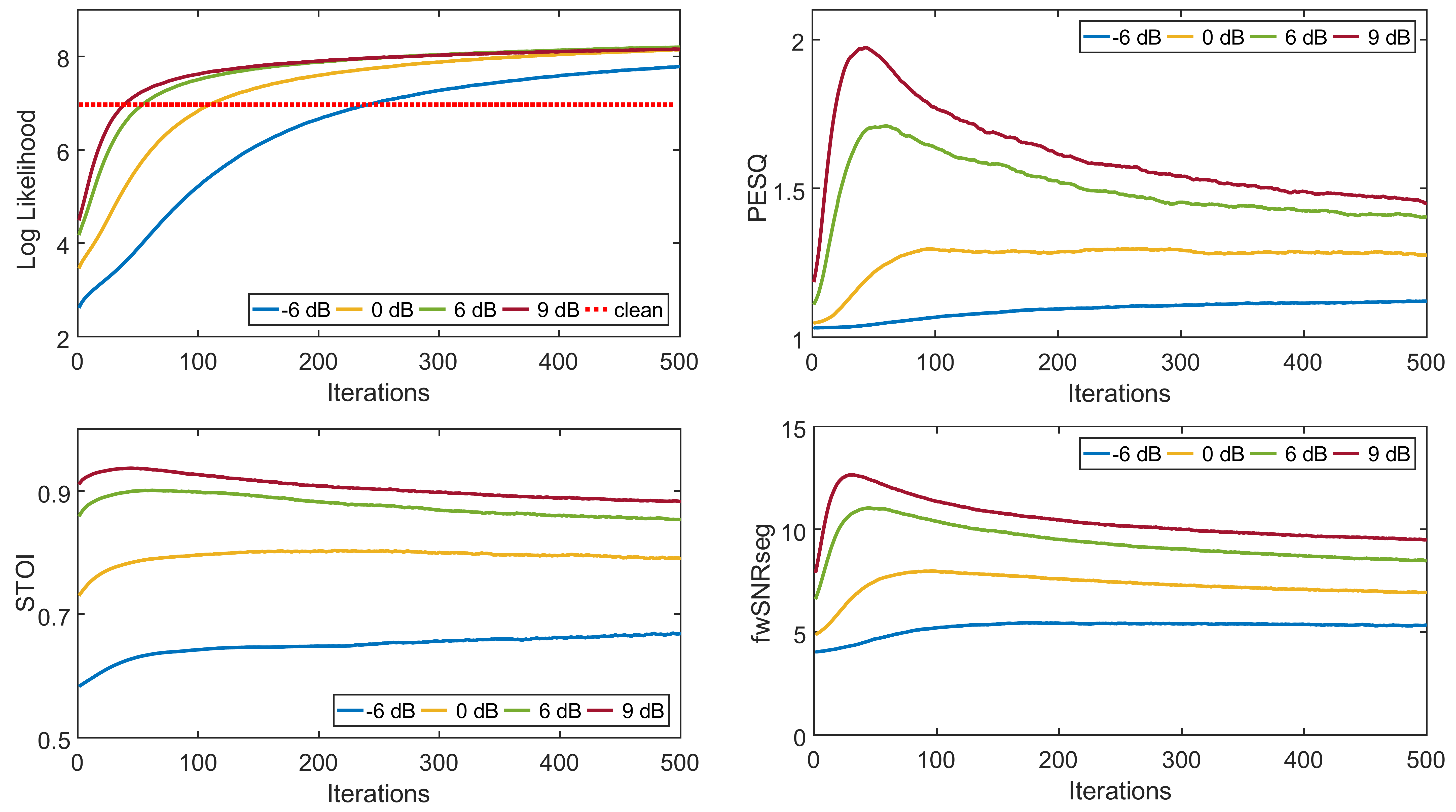}
	\caption{The change of log likelihood, PESQ, STOI and fwSNRseg during the inference, based on the Flow-Gauss (Opt) model.}
	\label{fig:trend-Flow-Gauss-opt}
\end{figure}

Although promising, we found that the Flow-Gauss model does not result in an optimal SE performance,
even with the oracle setting for noise.
To see this, we randomly selected 30 utterances from the training set and corrupted them by noise.
With the Flow-Gauss (Opt) setting, the optimization is conducted for 500 iterations, and
the log likelihood, PESQ, STOI and fwSNRseg are shown in Fig.~\ref{fig:trend-Flow-Gauss-opt}.

It can be found that the likelihoods in different SNR conditions are constantly increased with more iterations,
even higher than the likelihood of clean speech in the training set (read dot line).
However, the three SE metrics show a different pattern: they are firstly increased (indicating a better
speech quality), and then are decreased with more iterations. This clearly demonstrated that
a higher likelihood does not mean a better speech quality.

\begin{figure}[h]
	\centering
	\includegraphics[width=\linewidth]{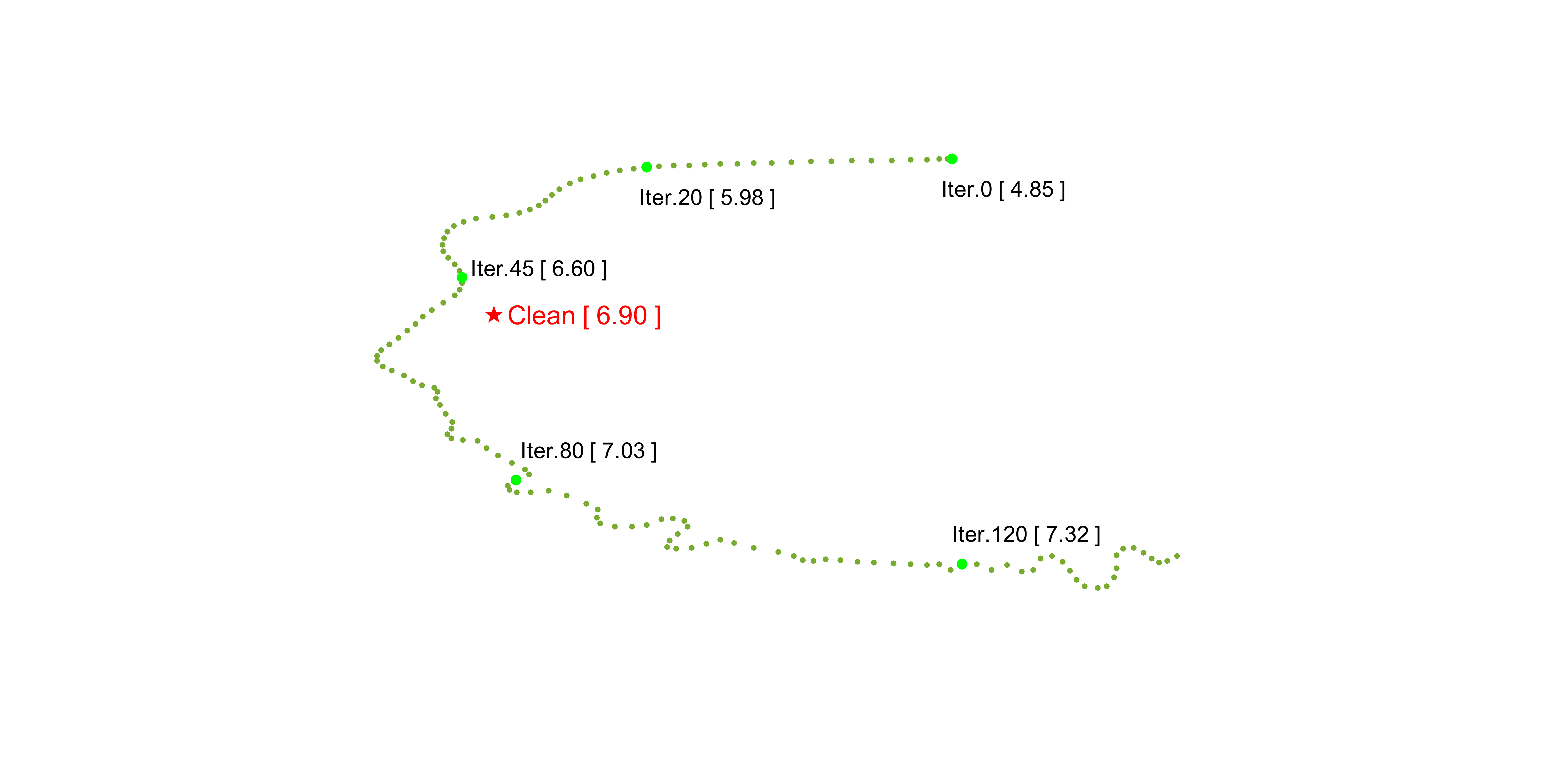}
	\caption{Footprint of the inference process with the Flow-Gauss (Opt) model, plot by t-SNE~\cite{tsne}. The red star is the target clean speech, and
    the green dots represent the enhanced speech at different steps of the optimization. [ \# ] is the log likelihood.}
	\label{fig:tracetsne}
\end{figure}

Fig.~\ref{fig:tracetsne} shows the optimization process for one noisy speech,
where the enhanced speech after a particular iteration is presented by a
green dot.
It can be observed that the optimization process attempted to approach the target clean speech,
however it just passed by the target and continued its journey to pursue a higher likelihood.

%It clearly demonstrated that the target clean speech is not the maximum likelihood.

%To observe the goal of the ML optimization, Fig.~\ref{fig:spec-post-opt}
%shows the spectrograms of enhanced speech outputs by Flow-Gauss (Opt) during the optimization process.
%It can be observed that the noise is gradually removed during the optimization process, while it does not stop when the noise has been
%mostly removed. It continuously removes the speech signal, in particular the high-frequency details.
%Again, this confirms that the expected optimal solution according to Eq.(\ref{eq:post}) does not really optimal in practice.

%\begin{figure}[h]
%	\centering
%	\includegraphics[width=\linewidth]{fig/fig4.pdf}
%	\caption{The enhanced speech outputs during the optimization process with Flow-Gauss (Opt).}
%	\label{fig:spec-post-opt}
%\end{figure}

\subsection{Analysis 2: Unreliable likelihood is caused by the deep prior model}

\begin{figure}[htp!]
  \centering
   \includegraphics[width=\linewidth]{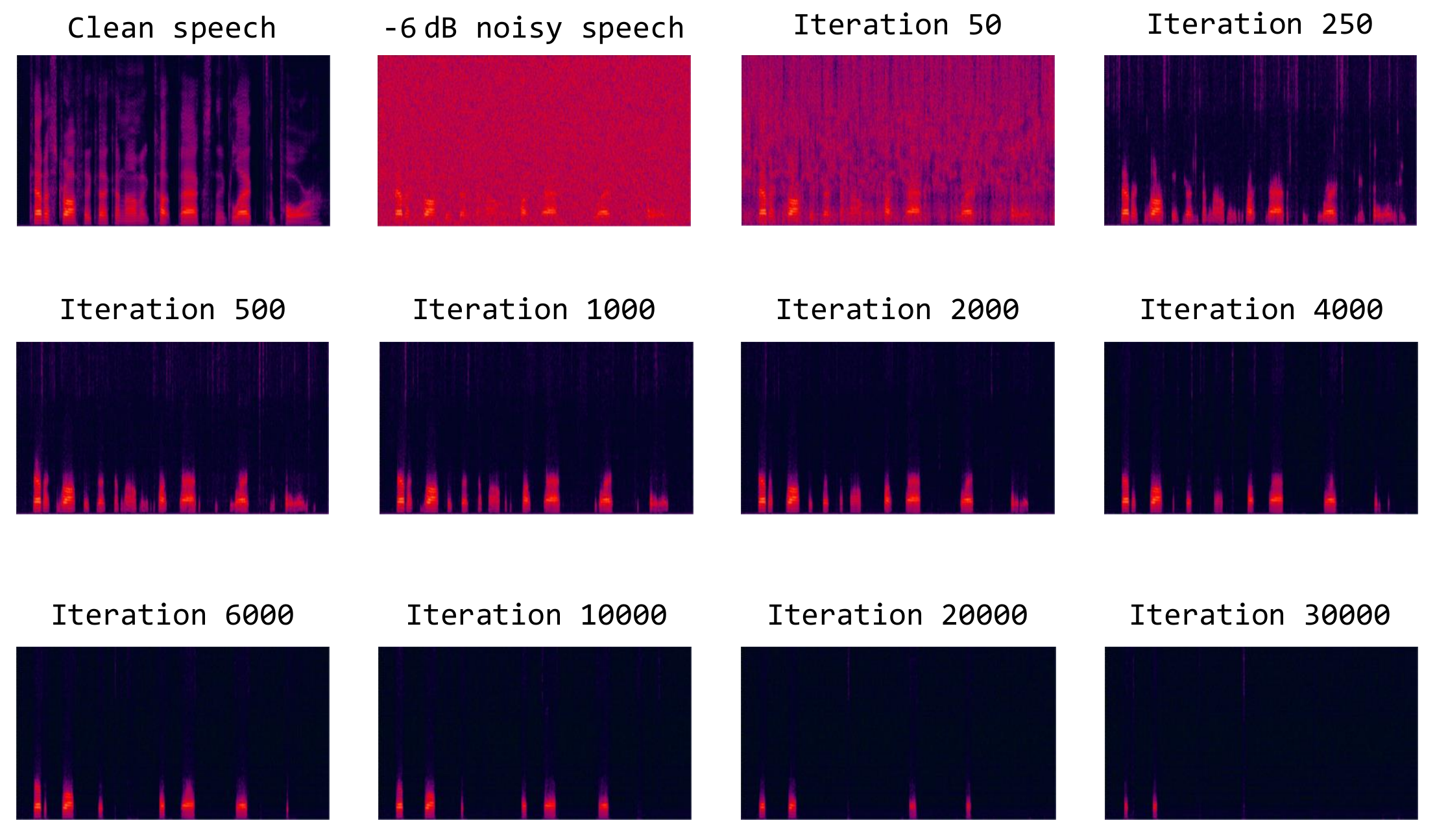}
   \caption{The enhanced speech from the optimization process that maximizes the prior likelihood $p(\mathbf{x})$.}
   \label{fig:spec-prior}
\end{figure}

Since the noise model is ideal and the ML inference is theoretically solid,
the failure of the ML-oriented optimization must be attributed to the deep speech prior model,
i.e., WaveGlow in our setting. Specifically, the model has assigned a high likelihood to an
unreal speech and so the ML optimization was attracted to a weird state, as shown in Fig.~\ref{fig:tracetsne}.
This argument can be easily verified by maximizing $p(\mathbf{x})$ provided by the prior model and see
where the optimization goes. The results are shown in Fig.~\ref{fig:spec-prior}.

It can be observed that the optimization indeed tries to remove the noise at the beginning, however
it does not stop when a reasonable enhancement has been obtained.
After thousands of iterations, the remaining spectrum tends to be zero.
It indicates that the speech prior assigns a higher likelihood to silence.
Notably, this observation coincides with the results in Fig.~\ref{fig:vae} where
the speech prior model is a VAE.

We argue that the unreliable likelihood associated with the deep speech prior model
is caused by the ML criterion used for model training. More specifically,
the goal of the ML training is to increase the probability density at the
locations of the training data.
To achieve this goal, the optimizer will try to twist the density function by using a complex transformation function.
However, this twisted density function is very risky and may assign unexpected high likelihood for
unreal data.

\begin{figure}[!htp]
	\centering
	\includegraphics[width=\linewidth]{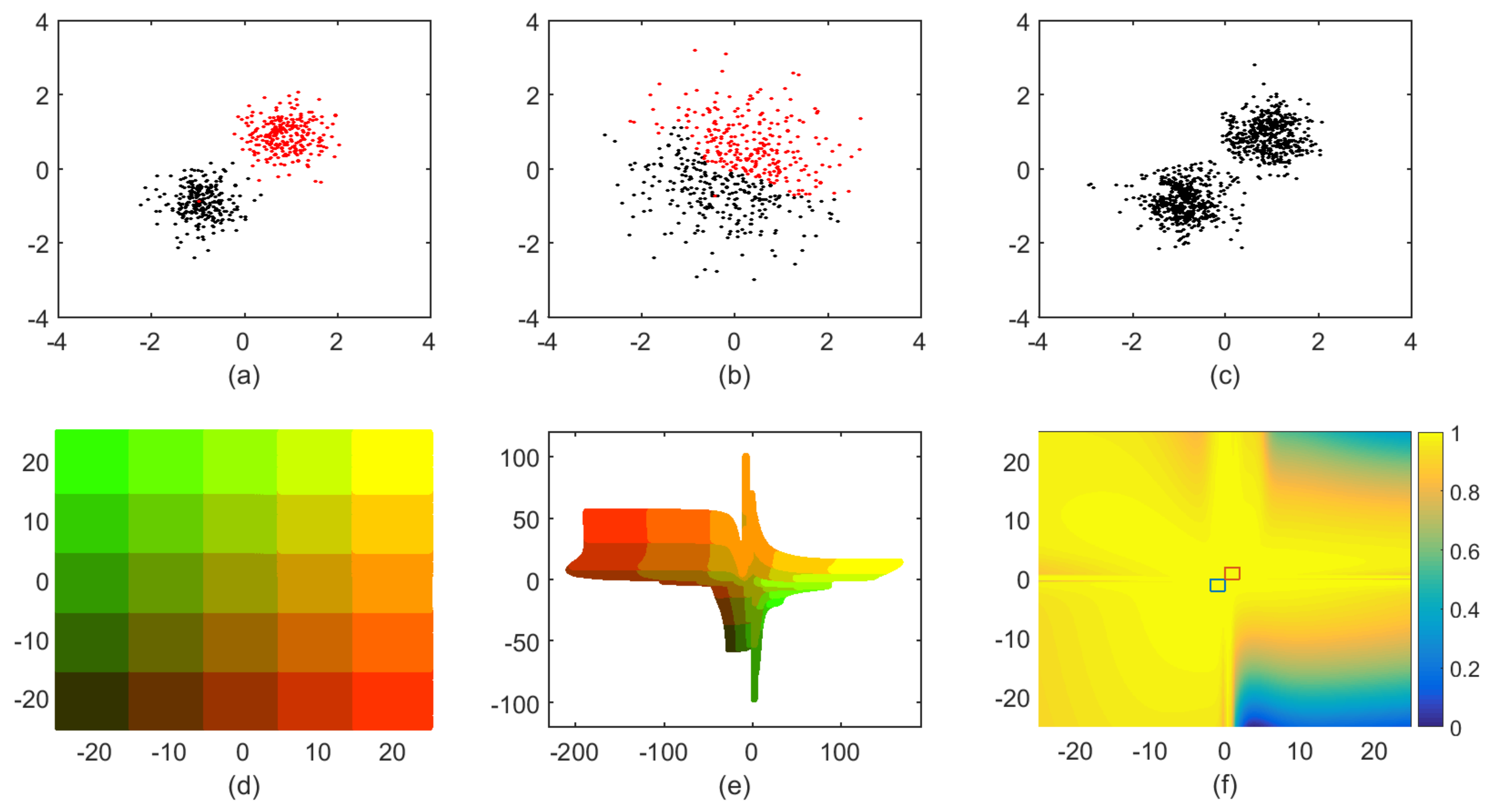}
	\caption{The density function produced by a 2-dimensional MAF model. The pictures are: (a) the training data; (b) the training data transformed into the latent space; (c) data
sampled from the model; (d) a color-grid in the data space; (e) the color-grid transformed to the latent space; (f) density function in the data space. The two squares in (f)
mark the two Gaussians of the training data shown in (a).}
	\label{fig:simu}
\end{figure}

A simulation experiment is conducted to verify this argument, where we use a flow model based on
the MAF structure~\cite{papamakarios2017masked}. MAF is much simpler than WaveGlow,
and can help us identify the problem by using a small set of simulation data. We sample 500 2-dimensional data
from two Gaussians, and train the MAF (5 blocks) for 2500 iterations. After the training,
the transformation and the density function are shown in Fig.~\ref{fig:simu}. It can be seen that
with such a simple flow, the transformation is highly complex and the density function is noticeably twisted.
Most importantly, there is a large region that is far from the training data but the density is high.

\section{Conclusion}

We conducted a comprehensive study for the behavior of deep speech prior models in speech enhancement.
To ensure a focused investigation, we choose normalization flow as the deep prior model,
build the model in the time domain, constrain the noise to be Gaussian, and use gradient descent to perform inference.
Our experiments show that our flow-based SE approach obtains comparable or even better performance than the
state-of-the-art VAE-NMF approach. However, further analyses show that the present result is far from optimal, due to the unreliable
likelihood produced by the flow model. We argue that this is a general problem for all deep generative models
based on ML training, and demonstrated that it may be caused by the inharmonious combination of the complex distribution transformation
and the ML-based training criterion. Therefore, caution must be given when employing such ML-based deep prior models
in speech enhancement.
More training data may alleviate this problem, but a principled solution requires more theoretical study.

\bibliographystyle{IEEEbib}
\bibliography{ref}

\end{document}